\documentclass[epj]{svjour}
\usepackage{amsfonts}
\usepackage{amsmath}
\usepackage{graphicx}
\usepackage{dcolumn}
\usepackage{bm}
\usepackage{booktabs}
\usepackage{feynmf}   
\usepackage{epstopdf}   
\usepackage{slashed}  
\usepackage{cases}
\usepackage{color}
\usepackage{multirow}
\usepackage{hyperref}
\usepackage{widetext}

\begin{document}

\title{QCD sum rule studies on the $s s \bar s \bar s$ tetraquark states of $J^{PC} = 0^{-+}$}

\author{Rui-Rui Dong\inst{1} \and Niu Su\inst{1} \and Hua-Xing Chen\inst{1,2} \and Er-Liang Cui\inst{3} \and Zhi-Yong Zhou\inst{2}
}                     
\offprints{}          
\institute{
School of Physics, Beihang University, Beijing 100191, China
\and
School of Physics, Southeast University, Nanjing 210094, China
\and
College of Science, Northwest A\&F University, Yangling 712100, China
}
\date{Received: date / Revised version: date}
%
\abstract{
We apply the method of QCD sum rules to study the $s s \bar s \bar s$ tetraquark states of $J^{PC} = 0^{-+}$. We construct all the relevant $s s \bar s \bar s$ tetraquark currents, and find that there are only two independent ones. We use them to further construct two weakly-correlated mixed currents. One of them leads to reliable QCD sum rule results and the mass is extracted to be $2.51^{+0.15}_{-0.12}$ GeV, suggesting that the $X(2370)$ or the $X(2500)$ can be explained as the $ss\bar s\bar s$ tetraquark state of $J^{PC} = 0^{-+}$. To verify this interpretation, we propose to further study the $\pi\pi/K \bar K$ invariant mass spectra of the $J/\psi \to \gamma \pi \pi \eta^\prime/\gamma K \bar K \eta^\prime$ decays in BESIII to examine whether there exists the $f_0(980)$ resonance.
} 
\maketitle

%
\section{Introduction}\label{sec:intro}
%

In the past twenty years there were a lot of exotic hadrons observed in particle experiments~\cite{pdg}, which can not be well explained in the traditional quark model~\cite{Liu:2019zoy,Lebed:2016hpi,Esposito:2016noz,Guo:2017jvc,Ali:2017jda,Olsen:2017bmm,Karliner:2017qhf,Brambilla:2019esw,Guo:2019twa}. Most of them contain one or two heavy quarks, and there are only a few exotic hadrons in the light sector composed only by up/down/$strange$ quarks. However, this situation is changing now. With a large amount of $J/\psi$ sample, the BESIII Collaboration are carefully examining the physics happening in the energy region around 2.0~GeV~\cite{Bai:2003sw,Ablikim:2005um,BESIII:2010krt,Ablikim:2010au,Ablikim:2016hlu,Ablikim:2019zyw,Ablikim:2020pgw}. Such experiments can also be performed by Belle-II~\cite{Kou:2018nap} and GlueX~\cite{Austregesilo:2018mno}, etc.

In Ref.~\cite{Ablikim:2010au}, the BESIII Collaboration observed two resonances $X(2120)$ and $X(2370)$ in the $\pi \pi \eta^\prime$ invariant mass spectrum of the $J/\psi \to \gamma \pi \pi \eta^\prime$ decay, together with the $X(1835)$~\cite{Bai:2003sw,Ablikim:2005um,BESIII:2010krt}. Recently in Ref.~\cite{Ablikim:2019zyw}, they further studied the $J/\psi \to \gamma K \bar K \eta^\prime$ decay, and observed the $X(2370)$ in the $K \bar K \eta^\prime$ invariant mass spectrum with a statistical significance of $8.3\sigma$, but they did not observed the $X(2120)$ in this process. This indicates that the $X(2370)$ probably contains many $strangeness$ components, more than the $X(2120)$. Besides, in Ref.~\cite{Ablikim:2016hlu}, they observed another resonance $X(2500)$ in the $\phi \phi$ invariant mass spectrum of the $J/\psi \to \gamma \phi \phi$ decay, which also contains many $strangeness$ components. The experimental parameters of the $X(2370)$ and $X(2500)$ were measured in these experiments to be:
\begin{eqnarray}
X(2370)     &:& M      = 2341.6 \pm 6.5 \pm 5.7~{\rm MeV}/c^2 \, ,
\\ \nonumber && \Gamma = 117 \pm 10 \pm 8~{\rm MeV} \, ,
\\ X(2500)  &:& M      = 2470\,^{+15}_{-19}\,^{+101}_{-23}~{\rm MeV}/c^2  \, ,
\\ \nonumber && \Gamma = 230\,^{+64}_{-35}\,^{+56}_{-33}~{\rm MeV} \, .
\end{eqnarray}

All these experimental observations inspire us to carefully investigate those hadrons containing many $strangeness$ components. One of the best candidates is the $s s \bar s \bar s$ tetraquark states, and the advantages to study them are: a) experimentally the widths of these resonances, if exist, are possibly not too broad, so they are capable of being observed; b) theoretically their internal structures are simpler than other multiquark states due to the Pauli principle restricting on identical $strangeness$ quarks, so their potential number is limited (this also makes them easier to be observed).

In this paper we shall study the $s s \bar s \bar s$ tetraquark states of $J^{PC} = 0^{-+}$ using the method of QCD sum rules. We have used the same approach in Refs.~\cite{Chen:2008ej,Chen:2018kuu,Cui:2019roq} to study the $s s \bar s \bar s$ tetraquark states of $J^{PC} = 1^{\pm-}$, where we found that there are only two independent $s s \bar s \bar s$ tetraquark currents of $J^{PC} = 1^{--}$ as well as two of $J^{PC} = 1^{+-}$.

Similarly, in the present study we shall find that there are only two independent $s s \bar s \bar s$ interpolating currents of $J^{PC} = 0^{-+}$. This makes it possible to perform a rather complete QCD sum rule analysis using both their diagonal and off-diagonal two-point correlation functions, from which we can further construct two weakly-correlated currents. We shall use them to perform QCD sum rule analyses, and the obtained results will be used to check whether the $X(2370)$ or the $X(2500)$ can be explained as the $s s \bar s \bar s$ tetraquark state of $J^{PC} = 0^{-+}$.

Before doing this, we note that the $s s \bar s \bar s$ tetraquark state is just one possibility, and there have been some other interpretations proposed to explain the $X(2370)$ and $X(2500)$. The $X(2370)$ is explained as
\begin{itemize}

\item a mixture of $\eta^\prime(4^1S_0)$ and glueball in Ref.~\cite{Liu:2010tr} within the framework of $^3P_0$ model (see also discussions in Ref.~\cite{Qin:2017qes});

\item the fourth radial excitation of $\eta(548)/\eta^\prime(958)$ in Ref.~\cite{Yu:2011ta} using the quark pair creation model;

\item a compact hexaquark state of $I^GJ^{PC} = 0^+0^{-+}$ in Ref.~\cite{Deng:2012wi} using the flux tube model;

\item a pseudoscalar glueball in Ref.~\cite{Eshraim:2012jv} based on a chirally invariant effective Lagrangian and in Ref.~\cite{Gui:2019dtm} using lattice QCD in quenched approximation.

\end{itemize}
The $X(2500)$ is explained as the $5^1S_0$ $s\bar s$ state using the $^3P_0$ model in Refs.~\cite{Pan:2016bac,Xue:2018jvi} and using the flux-tube model in Ref.~\cite{Wang:2017iai}. More Lattice QCD studies can be found in Refs.~\cite{Morningstar:1999rf,Chen:2005mg,Richards:2010ck,Gregory:2012hu,Eshraim:2016mds,Eshraim:2019sgr}, and their relevant dynamical analyses 
can be found in Refs.~\cite{Napsuciale:2007wp,MartinezTorres:2008gy,Liang:2013yta,Kozhevnikov:2019lmy,Lebiedowicz:2019jru,Kozhevnikov:2019rma}.

This paper is organized as follows. In Sec.~\ref{sec:current}, we systematically construct the $s s \bar s \bar s$ tetraquark currents of $J^{PC} = 0^{-+}$, and find two independent currents $\eta_1$ and $\eta_2$. We use them to perform QCD sum rule analyses in Sec.~\ref{sec:sumrule}, and calculate both their diagonal and off-diagonal two-point correlation functions. Then we perform numerical analyses using the two single currents $\eta_1$ and $\eta_2$ in Sec.~\ref{sec:single}, and using the two weakly-correlated mixed currents $J_1$ and $J_2$ in Sec.~\ref{sec:mixing}. Sec.~\ref{sec:summary} is a summary.

%
\section{Interpolating Currents}\label{sec:current}
%

In this section we construct the $s s \bar s \bar s$ tetraquark currents with the spin-parity quantum number $J^{PC} = 0^{-+}$. There are two non-vanishing diquark-antidiquark currents:
%
\begin{eqnarray}
\eta_{1} &=& (s_a^T C s_b) (\bar{s}_a \gamma_5 C \bar{s}_b^T)
\label{def:eta1} + (s_a^T C \gamma_5 s_b) (\bar{s}_a C \bar{s}_b^T)  \, ,
\\ \eta_{2} &=& (s_a^T C \sigma_{\mu\nu} s_b) (\bar{s}_a \sigma^{\mu\nu} \gamma_5 C \bar{s}_b^T) \, .
\label{def:eta2}
\end{eqnarray}
%
In the above expressions $a$ and $b$ are color indices, and the sum over repeated indices is taken. These two currents are independent of each other.

Since the diquark fields $s_a^T C s_b/s_a^T C \gamma_5 s_b/s_a^T C \sigma_{\mu\nu} s_b/s_a^T C \sigma_{\mu\nu} \gamma_5 s_b$ have the quantum numbers $J^P = 0^-/0^+/1^\pm/1^\mp$ respectively, the former current $\eta_{1}$ contains one purely ground-state diquark/antidiquark field and one purely excited one, while the latter $\eta_{2}$ contains two ``partially-ground-state-partially-excited'' diquark/antidiquark fields. Besides, the former current $\eta_1$ has the symmetric color structure $(ss)_{\mathbf{6}_C}(\bar s \bar s)_{\mathbf{\bar 6}_C}$, while the latter $\eta_2$ has the antisymmetric color structure $(ss)_{\mathbf{\bar 3}_C}(\bar s \bar s)_{\mathbf{3}_C}$. Hence, it is not easy to tell at this moment which one has a more stable internal structure and leads to better sum rule results.

Besides $\eta_{1}$ and $\eta_{2}$, we can construct four mesonic-mesonic currents:
%
\begin{eqnarray}
\eta_{3} &=& (\bar{s}_a s_a)(\bar{s}_b \gamma_5 s_b) \, ,
\\ \eta_{4} &=& (\bar{s}_a \sigma_{\mu\nu} s_a)(\bar{s}_b \sigma^{\mu\nu} \gamma_5 s_b) \, ,
\\ \eta_{5} &=& {\lambda_{ab}}{\lambda_{cd}}(\bar{s}_a s_b)(\bar{s}_c \gamma_5 s_d) \, ,
\\ \eta_{6} &=& {\lambda_{ab}}{\lambda_{cd}} (\bar{s}_a \sigma_{\mu\nu} s_b)(\bar{s}_c \sigma^{\mu\nu} \gamma_5 s_d) \, .
\end{eqnarray}
%
The former two $\eta_{3,4}$ have the color structure $(\bar s s)_{\mathbf{1}_c}(\bar s s)_{\mathbf{1}_c}$, and the latter two $\eta_{5,6}$ have the color structure $(\bar s s)_{\mathbf{8}_c}(\bar s s)_{\mathbf{8}_c}$. However, only two of them are independent due to the following relations derived using the Fierz transformation:
%
\begin{eqnarray}
\eta_{5} &=& - \frac{5}{3}~\eta_{3} - {1\over4}~\eta_{4} \, ,
\label{eq:fierz1}
\\ \nonumber \eta_{6} &=& -12~\eta_{3} + \frac{1}{3}~\eta_{4} \, .
\end{eqnarray}
%
Moreover, we can apply the Fierz transformation to extract the following relations between diquark-antidiquark and mesonic-mesonic currents:
%
\begin{eqnarray}
\eta_{1} &=& - \eta_{3} + {1\over4}~\eta_{4} \, ,
\label{eq:fierz2}
\\ \nonumber \eta_{2} &=& 6~\eta_{3} - {1\over2}~\eta_{4} \, .
\end{eqnarray}
%
Therefore, these two constructions are equivalent. We shall use these identities to investigate decay properties at the end of this paper.

In the following we shall use $\eta_{1}$ and $\eta_{2}$ to perform QCD sum rule analyses, and separately calculate their diagonal two-point correlation functions:
\begin{eqnarray}
\Pi_{11}(x) &=& \langle 0 | {\bf T}[ \eta_{1}(x) { \eta_{1}^\dagger } (0)] | 0 \rangle \, ,
\\ \nonumber \Pi_{22}(x) &=& \langle 0 | {\bf T}[ \eta_{2}(x) { \eta_{2}^\dagger } (0)] | 0 \rangle \, .
\end{eqnarray}
Moreover, we shall calculate their off-diagonal term:
\begin{equation}
\Pi_{12}(x) = \langle 0 | {\bf T}[ \eta_{1}(x) { \eta_{2}^\dagger } (0)] | 0 \rangle \, ,
\label{eq:offdiag1}
\end{equation}
and we shall find that these two currents strongly correlate with each other.

Based on the above diagonal and off-diagonal correlation functions, we shall further construct two weakly-correlated currents
\begin{eqnarray}
J_{1} &=& \cos\theta~\eta_{1} + \sin\theta~\eta_{2} \, ,
\\ \nonumber J_{2} &=& - \sin\theta~\eta_{1} + \cos\theta~\eta_{2} \, .
\end{eqnarray}
After choosing a suitable mixing angle, we shall find them to satisfy:
\begin{eqnarray}
&& \langle 0 | {\bf T}[ J_{1}(x) J_{2}^\dagger (0)] | 0 \rangle
\label{eq:offdiag2}
\\ \nonumber &\ll& \left(\langle 0 | {\bf T}[ J_{1}(x) { J_{1}^\dagger } (0)] | 0 \rangle
\times \langle 0 | {\bf T}[ J_{2}(x) { J_{2}^\dagger } (0)] | 0 \rangle\right)^{1/2} \, ,
\end{eqnarray}
in proper working regions. In the following we shall also use $J_{1}$ and $J_{2}$ to perform QCD sum rule analyses.

%
\section{QCD sum rule Analysis}
\label{sec:sumrule}
%

In the method of QCD sum rules~\cite{Shifman:1978bx,Reinders:1984sr} one needs to calculate the two-point correlation function
%
\begin{equation}
\Pi(q^2) \equiv i \int d^4x e^{iqx} \langle 0 | {\bf T}[ \eta(x) { \eta^\dagger } (0)] | 0 \rangle \, ,
\label{def:pi}
\end{equation}
%
at both hadron and quark-gluon levels.

Firstly, at the hadron level we express Eq.~(\ref{def:pi}) using the dispersion relation:
%
\begin{equation}
\Pi(q^2) = \int^\infty_{s_<}\frac{\rho(s)}{s-q^2-i\varepsilon}ds \, ,
\end{equation}
%
where $s_<$ denotes the physical threshold, and it is $s_< = 16 m_s^2$ in the present case; $\rho(s)$ is the spectral density, parameterized using one
pole dominance for the ground state $X$ together with a continuum contribution:
%
\begin{eqnarray}
\rho(s) &\equiv& \sum_n\delta(s-M^2_n) \langle 0| \eta | n\rangle \langle n| {\eta^\dagger} |0 \rangle
\\ \nonumber &=& f^2_X \delta(s-M^2_X) + \rm{continuum} \, .
\label{eq:rho}
\end{eqnarray}
%
Here $f_X$ is the decay constant, defined as
\begin{eqnarray}
\langle 0 | \eta | X \rangle &=& f_X \, .
\end{eqnarray}

Secondly, at the quark-gluon level we insert $\eta_{1}$ and $\eta_{2}$ into Eq.~(\ref{def:pi}) and calculate it using the method of operator product expansion (OPE).

Thirdly, we perform the Borel transformation at both hadron and quark-gluon levels:
%
\begin{equation}
\Pi(M_B^2)\equiv\mathcal{B}_{M_B^2}\Pi(p^2)=\int^\infty_{s_<} e^{-s/M_B^2} \rho(s)ds \, .
\label{eq_borel}
\end{equation}
%
After approximating the continuum using the spectral density above a threshold value $s_0$, we obtain the sum rule equation
%
\begin{equation}
\Pi(s_0, M_B^2) \equiv f^2_X e^{-M_X^2/M_B^2} = \int^{s_0}_{s_<} e^{-s/M_B^2}\rho(s)ds \, ,
\label{eq_fin}
\end{equation}
%
which can be used to calculate $M_X$ through
%
\begin{eqnarray}
M^2_X(s_0, M_B) &=& \frac{\frac{\partial}{\partial(-1/M_B^2)}\Pi(s_0, M_B^2)}{\Pi(s_0, M_B^2)}
\label{eq:LSR}
\\ \nonumber &=& \frac{\int^{s_0}_{s_<} e^{-s/M_B^2}s\rho(s)ds}{\int^{s_0}_{s_<} e^{-s/M_B^2}\rho(s)ds} \, .
\end{eqnarray}
%

%
\begin{widetext}
In the present study we calculate OPEs up to the $D({\rm imension}) = 10$ terms, including the perturbative term, the strange quark mass, the quark condensate, the gluon condensate, the quark-gluon mixed condensate, as well as their combinations:
\begin{eqnarray}
\nonumber \Pi_{11} &=& \int^{s_0}_{s_<} \Bigg [
{s^4 \over 15360 \pi^6}
-  {m_s^2\over192\pi^6} s^3
+ \Big (- {\langle g_s^2 GG \rangle\over3072\pi^6 }
+ {5 m_s^4\over64\pi^6}
+ {m_s\langle \bar s s \rangle\over24\pi^4}\Big )s^2
+ \Big({m_s^2\langle g_s^2 GG \rangle\over256\pi^6 }
- {3m_s^6\over8\pi^6 }
- {m_s^3\langle \bar s s \rangle\over4\pi^4}\Big)s
\\ \nonumber &&
+\Big ( -{m_s^4 \langle g_s^2 GG \rangle  \over 256 \pi^6 }
- { m_s \langle g_s^2 GG \rangle \langle \bar s s \rangle \over 192 \pi^4}
+ { 3 m_s^8  \over 16 \pi^6 }
+ { 3 m_s^5\langle \bar s s \rangle \over 2 \pi^4 }
+ {  m_s^3\langle g_s \bar s \sigma G s \rangle \over 4 \pi^4 }
- { 3 m_s^2\langle \bar s s \rangle^2 \over 2 \pi^2 }\Big ) \Bigg ] e^{-s/M_B^2} ds
\\ && \label{eq:pieta1}
+ \Big ( {m_s^3 \langle g_s^2 GG \rangle  \langle \bar s s \rangle \over 192 \pi^4}
- { m_s^7 \langle \bar s s \rangle \over 2 \pi^4 }
- { m_s^4 \langle \bar s s \rangle^2 \over \pi^2}
- {m_s^2 \langle \bar s s \rangle  \langle g_s \bar s \sigma G s \rangle \over \pi^2 }
+ { 16 m_s \langle \bar s s \rangle^3 \over 9 } \Big )
\, ,
\\ \nonumber
\Pi_{22} &=& \int^{s_0}_{s_<} \Bigg [
{s^4 \over 2560 \pi^6}
 -{ m_s^2  \over 32 \pi^6 }  s^3
+ \Big ( {\langle g_s^2 GG \rangle \over 768 \pi^6 }
+ {15 m_s^4 \over 32 \pi^6}+{m_s \langle \bar s s \rangle \over 4 \pi^4 } \Big ) s^2
+ \Big ( -{m_s^2 \langle g_s^2 GG \rangle  \over 64 \pi^6 }
- {9 m_s^6  \over 4 \pi^6 }
- { 3 m_s^3 \langle \bar s s \rangle \over 2 \pi^4 } \Big ) s
\\ \nonumber &&
+ \Big ( {  m_s^4 \langle g_s^2 GG \rangle  \over 64 \pi^6}
+ { 9 m_s^8 \over 8 \pi^6 }
+ {m_s \langle g_s^2 GG \rangle \langle \bar s s \rangle \over 48 \pi^4}
 +{9 m_s^5 \langle \bar s s \rangle \over \pi^4 }
- {9 m_s^2\langle \bar s s \rangle^2 \over \pi^2 }+{ 3 m_s^3 \langle g_s \bar s \sigma G s \rangle \over 2 \pi^4 } \Big ) \Bigg ] e^{-s/M_B^2} ds
\\  \label{eq:pieta2} &&
+ \Big ( - {m_s^3 \langle g_s^2 GG \rangle \langle \bar s s \rangle\over 48 \pi^4 }
- { 3 m_s^7\langle \bar s s \rangle \over \pi^4 }
- { 6 m_s^4\langle \bar s s \rangle^2 \over \pi^2  }
- { 6 m_s^2\langle \bar s s \rangle \langle g_s \bar s \sigma G s \rangle \over \pi^2 }
+ { 32 m_s\langle \bar s s \rangle^3 \over 3 } \Big )
\, ,
\\ \label{eq:eta12}
\Pi_{12} &=& \int^{s_0}_{s_<} \Bigg [
- { \langle g_s^2 GG \rangle \over 512 \pi^6 }  s^2
+  { 3 m_s^2 \langle g_s^2 GG \rangle  \over 128 \pi^6 }  s
- { 3 m_s^4 \langle g_s^2 GG \rangle  \over 128 \pi^6}
- {m_s\langle g_s^2 GG \rangle  \langle \bar s s \rangle \over 32 \pi^4}\Bigg ] e^{-s/M_B^2} ds
+ { m_s^3\langle g_s^2 GG \rangle \langle \bar s s \rangle \over 32 \pi^4}
\, .
\end{eqnarray}
\end{widetext}
Based on these expressions, we shall use the two single currents $\eta_{1}$ and $\eta_{2}$ to perform QCD sum rule analyses in Sec.~\ref{sec:single}, and use the two mixed currents $J_{1}$ and $J_{2}$ to perform QCD sum rule analyses in Sec.~\ref{sec:mixing}. In the calculations we shall use the following values for various quark and gluon parameters~\cite{Yang:1993bp,Narison:2002pw,Gimenez:2005nt,Jamin:2002ev,Ioffe:2002be,Ovchinnikov:1988gk,Ellis:1996xc,pdg}:
%
\begin{eqnarray}
\nonumber m_s(2\mbox{ GeV}) &=& 96 ^{+8}_{-4} \mbox{ MeV}\, ,
\\ \nonumber \langle\bar ss\rangle &=& -(0.8\pm 0.1)\times(0.240 \mbox{ GeV})^3\, ,
\\ \langle g_s^2GG\rangle &=& (0.48\pm 0.14) \mbox{ GeV}^4\, ,
\label{condensates}
\\ \nonumber \langle g_s\bar s\sigma G s\rangle &=& -M_0^2\times\langle\bar ss\rangle\, ,
\\ \nonumber M_0^2 &=& (0.8 \pm 0.2) \mbox{ GeV}^2\, .
\end{eqnarray}
%

%
\section{Single Currents $\eta_{1}$ and $\eta_{2}$}
\label{sec:single}
%

In this section we use the two single currents $\eta_{1}$ and $\eta_{2}$ to perform QCD sum rule analyses. When applying QCD sum rules to study multiquark states, one usually meets a serious problem, {\it i.e.}, how to differentiate the multiquark state and the relevant threshold, because the current may couple to both of them. In the present study the relevant threshold is the $\eta^\prime f_0(980)$ around 1950~MeV. Besides, $\eta_{1}$ and $\eta_{2}$ may also couple to the lower states of $J^{PC} = 0^{-+}$, such as the $\eta(1475)$, etc.

\begin{figure}[hbt]
\begin{center}
\includegraphics[width=0.47\textwidth]{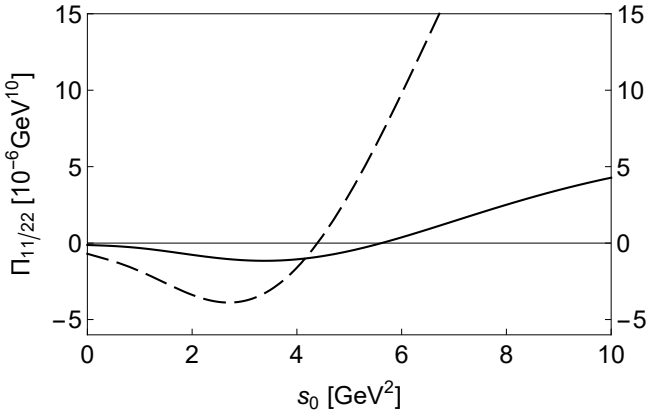}
\caption{The two-point correlation functions, $\Pi_{11}(s_0, M_B^2)$ (solid) and $\Pi_{22}(s_0, M_B^2)$ (dashed), as functions of the threshold value $s_0$. These curves are obtained by setting $M_B^2 = 1.5$~GeV$^2$.}
\label{fig:pi}
\end{center}
\end{figure}

If this happens, the resulting correlation function should be positive. However, as shown in Fig.~\ref{fig:pi}, we find that the two correlation functions $\Pi_{11}(M_B^2)$ and $\Pi_{22}(M_B^2)$ are both negative in the region $s_0 < 4.0$ GeV$^2$ when taking $M_B^2 = 1.5$~GeV$^2$. This fortunately indicates that both $\eta_{1}$ and $\eta_{2}$ do not strongly couple to the $\eta^\prime f_0(980)$ threshold as well as the lower state $\eta(1475)$. Hence, the state they couple to, as if they can couple to some state, should be new and possibly exotic. To investigate this state, the proper $s_0$ should be significantly larger than $4.0/6.0$ GeV$^2$, where $\Pi_{11}(M_B^2)/\Pi_{22}(M_B^2)$ are positive.

%
%

\begin{figure}[hbt]
\begin{center}
\includegraphics[width=0.47\textwidth]{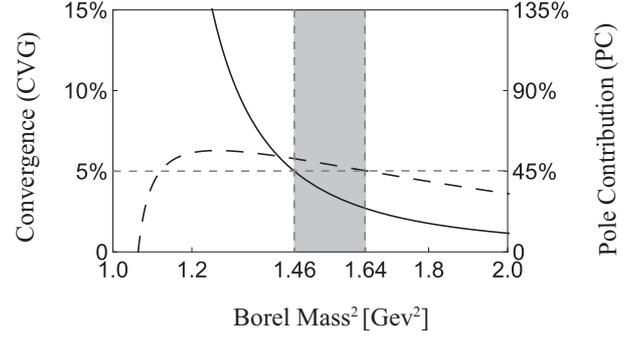}
\caption{CVG (solid curve, defined in Eq.~(\ref{eq:convergence})) and PC (dashed curve, defined in Eq.~(\ref{eq:pole})) as functions of the Borel mass $M_B$. These curves are obtained using the single current $\eta_{1}$ when setting $s_0 = 8.8$~GeV$^2$.}
\label{fig:cvgpole}
\end{center}
\end{figure}

To extract the mass of this exotic state, $M_X$, through Eq.~(\ref{eq:LSR}), we need to find proper working regions for the two free parameters, the threshold value $s_0$ and the Borel mass $M_B$. Taking $\eta_{1}$ as an example, first we investigate the convergence of the operator product expansion (CVG) by requiring the $D=10$ terms to be less than 5\%:
\begin{eqnarray}
\mbox{CVG} &\equiv& \left|\frac{ \Pi^{D=10}(s_0, M_B^2) }{ \Pi(s_0, M_B^2) }\right| \leq 5\% \, .
\label{eq:convergence}
\end{eqnarray}
This is the cornerstone of a reliable QCD sum rule analysis. As shown in Fig.~\ref{fig:cvgpole} using the solid curve, this condition is satisfied in the region $M_B^2 > 1.46$~GeV$^2$ when setting $s_0 = 8.8$~GeV$^2$.

%
%

Then we investigate the one-pole-dominance assumption by requiring the pole contribution (PC) to be larger than 45\%:
\begin{eqnarray}
\mbox{PC} &\equiv& \left|\frac{ \Pi(s_0, M_B^2) }{ \Pi(\infty, M_B^2) }\right| \geq 45\% \, ,
\label{eq:pole}
\end{eqnarray}
so that its average value is about 50\%. As shown in Fig.~\ref{fig:cvgpole} using the dashed curve, this condition is satisfied in the region $M_B^2 < 1.64$~GeV$^2$ when setting $s_0 = 8.8$~GeV$^2$. Altogether we obtain a Borel window $1.46$~GeV$^2 < M_B^2 < 1.64$~GeV$^2$ when setting $s_0 = 8.8$~GeV$^2$. We change $s_0$ to redo the same procedures, and find that there exist non-vanishing Borel windows as long as $s_0 \geq 8.4$~GeV$^2$.

\begin{figure*}[]
\begin{center}
\includegraphics[width=0.45\textwidth]{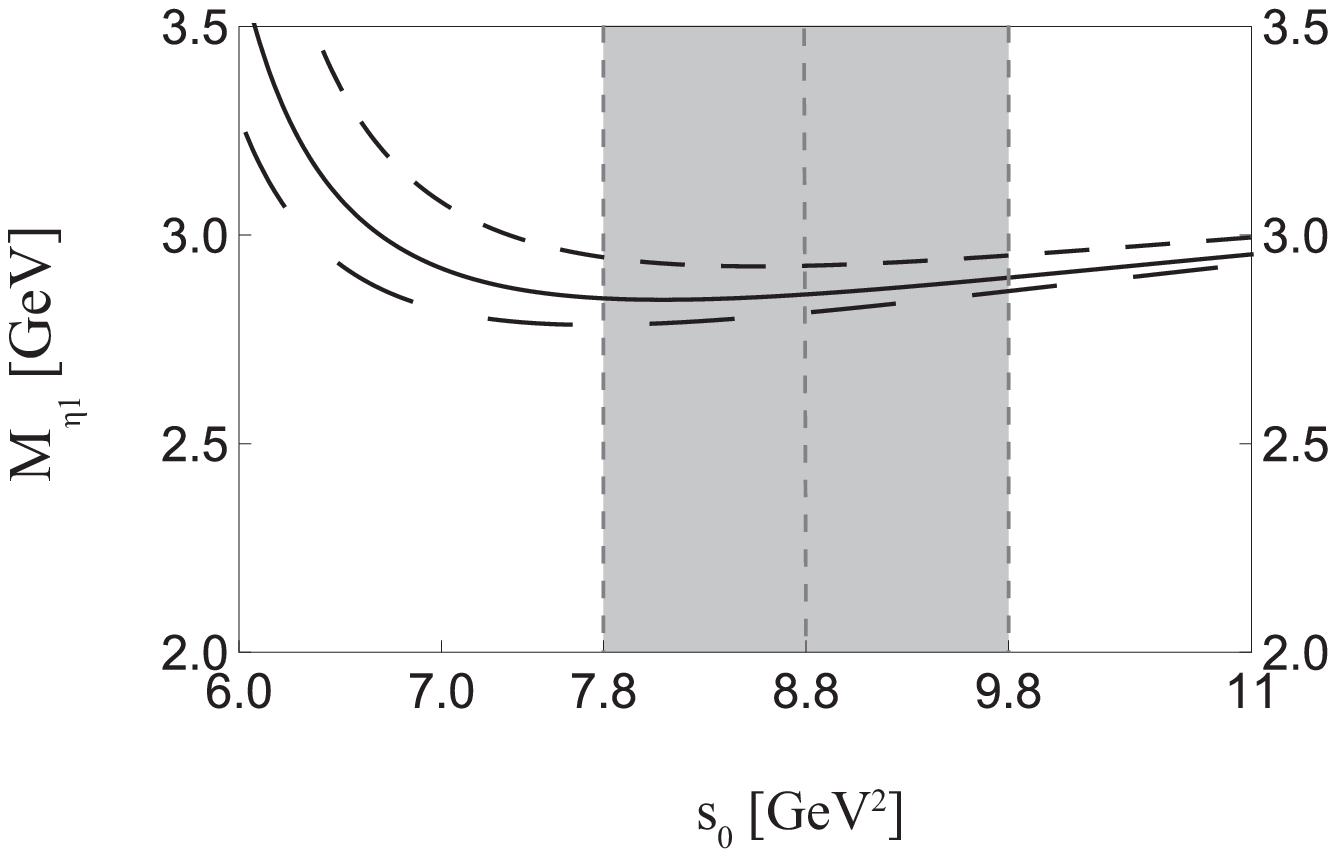}
~~~~~
\includegraphics[width=0.45\textwidth]{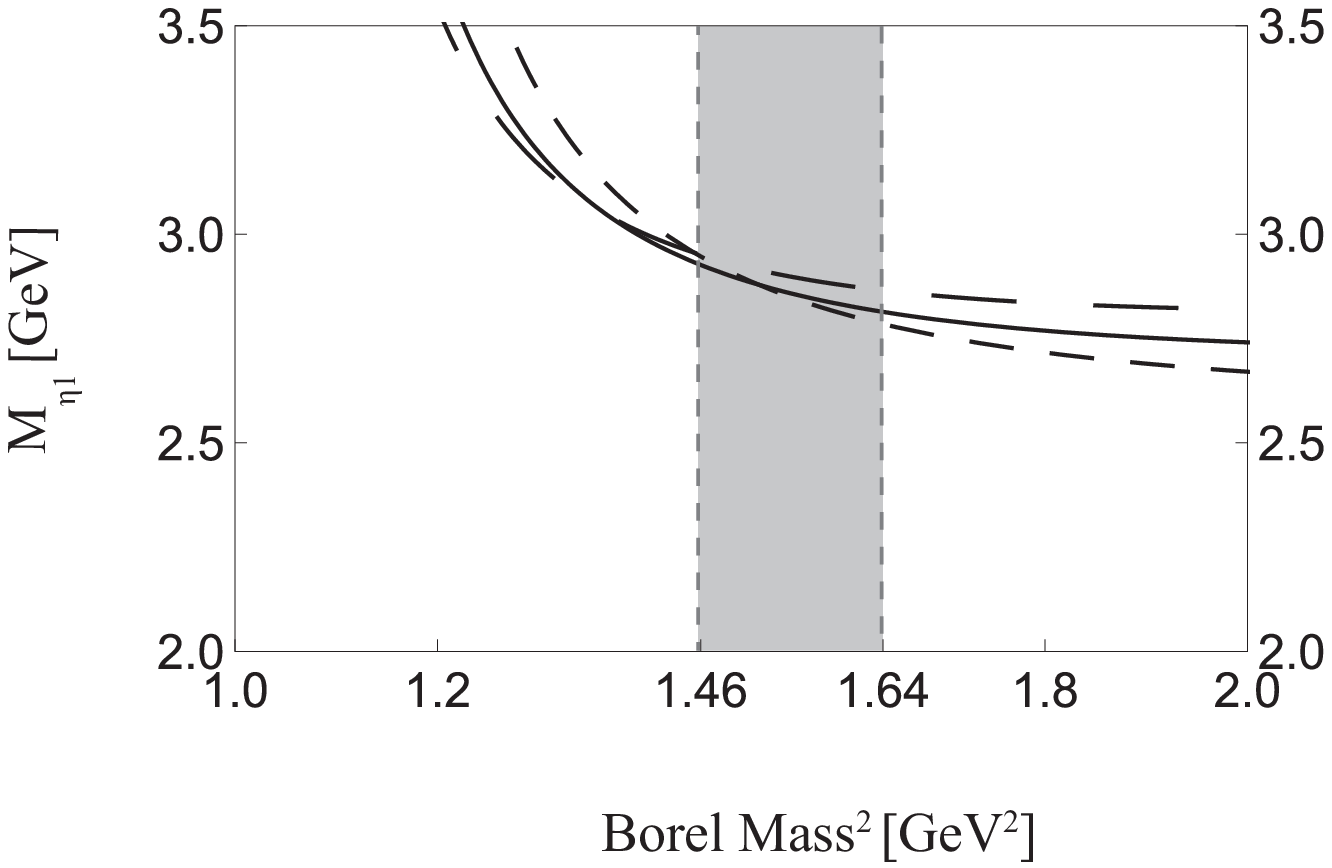}
\caption{
Mass calculated using the current $\eta_{1}$, as a function of the threshold value $s_0$ (left) and the Borel mass $M_B$ (right).
In the left panel the short-dashed/solid/long-dashed curves are obtained by setting $M_B^2 = 1.46/1.55/1.64$ GeV$^2$, respectively.
In the right panel the short-dashed/solid/long-dashed/dotted curves are obtained by setting $s_0 = 7.8/8.8/9.8$ GeV$^2$, respectively.}
\label{fig:eta1mass}
\end{center}
\end{figure*}

Finally, we require the mass $M_X$ extracted from Eq.~(\ref{eq:LSR}) to have a dual minimum dependence on both the threshold value $s_0$ and the Borel mass $M_B$. Still taking $\eta_{1}$ as an example, we show the mass $M_X$ in Fig.~\ref{fig:eta1mass} as a function of the threshold value $s_0$ (left) and the Borel mass $M_B$ (right). We find $M_X$ has a minimum around $s_0 \sim 8.8$~GeV$^2$, and its dependence on $M_B$ is moderate in the Borel window $1.46$~GeV$^2 < M_B^2 < 1.64$~GeV$^2$. Accordingly, we choose the working regions to be $7.8$~GeV$^2< s_0 < 9.8$~GeV$^2$ and $1.46$~GeV$^2 < M_B^2 < 1.64$~GeV$^2$, where the mass $M_X$ is evaluated to be
\begin{equation}
M_{\eta_1} = 2.86^{+0.18}_{-0.12}{\rm~GeV} \, .
\end{equation}
Here the central value corresponds to $s_0=8.8$~GeV$^2$ and $M_B^2 = 1.55$~GeV$^2$, and the uncertainty is due to the Borel mass $M_B$ and the threshold value $s_0$ as well as various quark and gluon parameters listed in Eqs.~(\ref{condensates}).

%
%

\begin{figure*}[]
\begin{center}
\includegraphics[width=0.45\textwidth]{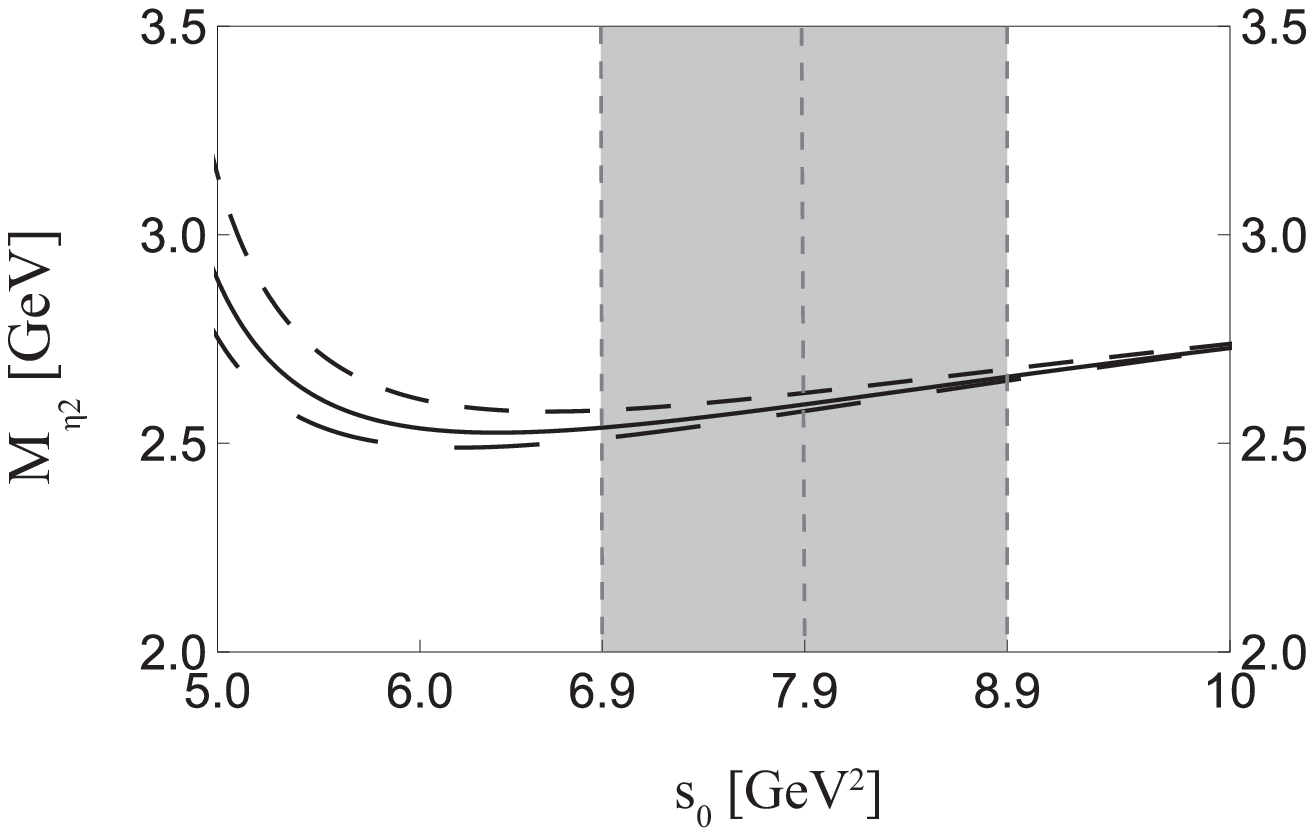}
~~~~~
\includegraphics[width=0.45\textwidth]{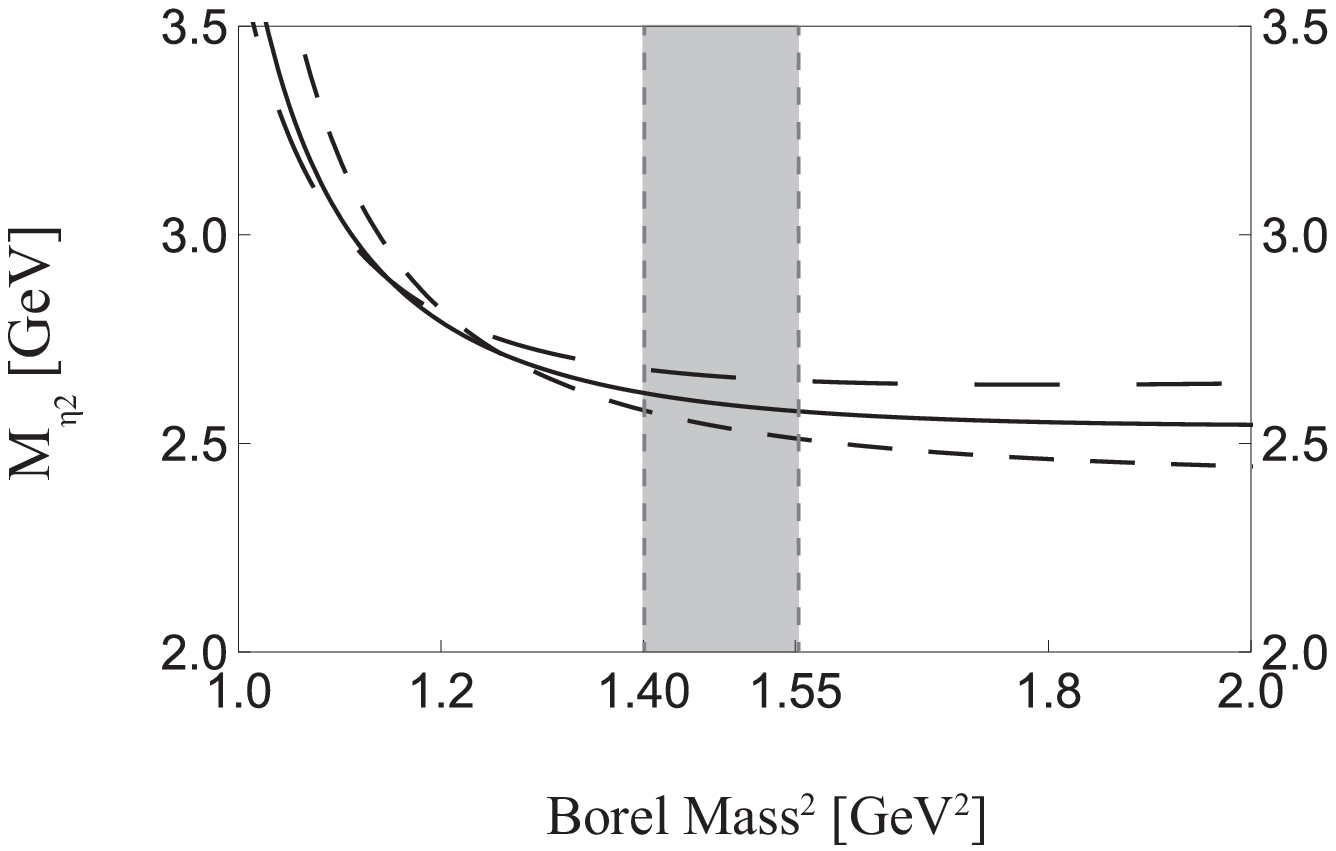}
\caption{
Mass calculated using the current $\eta_{2}$, as a function of the threshold value $s_0$ (left) and the Borel mass $M_B$ (right).
In the left panel the short-dashed/solid/long-dashed curves are obtained by setting $M_B^2 = 1.40/1.47/1.55$ GeV$^2$, respectively.
In the right panel the short-dashed/solid/long-dashed/dotted curves are obtained by setting $s_0 = 6.9/7.9/8.9$ GeV$^2$, respectively.}
\label{fig:eta2mass}
\end{center}
\end{figure*}

Similarly, we use $\eta_{2}$ to perform QCD sum rule analyses, and find that there exist non-vanishing Borel windows as long as $s_0 \geq 7.5$~GeV$^2$. Using the working regions $6.9$~GeV$^2< s_0 < 8.9$~GeV$^2$ and $1.40$~GeV$^2 < M_B^2 < 1.55$~GeV$^2$ (Borel window for $s_0=7.9$~GeV$^2$), we obtain
\begin{equation}
M_{\eta_2} = 2.59^{+0.14}_{-0.10}{\rm~GeV} \, ,
\end{equation}
where the central value corresponds to $s_0=7.9$~GeV$^2$ and $M_B^2 = 1.47$~GeV$^2$. For completeness, we show the mass obtained using $\eta_2$ in Fig.~\ref{fig:eta2mass} as a function of the threshold value $s_0$ (left) and the Borel mass $M_B$ (right). The mass dependence on $M_B$ is weak and acceptable in the Borel window $1.40$~GeV$^2 < M_B^2 < 1.55$~GeV$^2$, which is slightly better than the previous result obtained using $\eta_1$.

%
%

%
\section{Mixed Currents $J_{1}$ and $J_{2}$}
\label{sec:mixing}
%

In the previous section we have used the two single currents $\eta_{1}$ and $\eta_{2}$ to perform QCD sum rule analyses. In this section we further study their mixing, and use the two mixed currents $J_{1}$ and $J_{2}$ to perform QCD sum rule analyses. We follow the procedures used in Refs.~\cite{Chen:2018kuu,Cui:2019roq} to do this, where the mixing of $s s \bar s \bar s$ tetraquark currents with $J^{PC} = 1^{\pm-}$ is carefully investigated.

\begin{figure}[hbt]
\begin{center}
\includegraphics[width=0.48\textwidth]{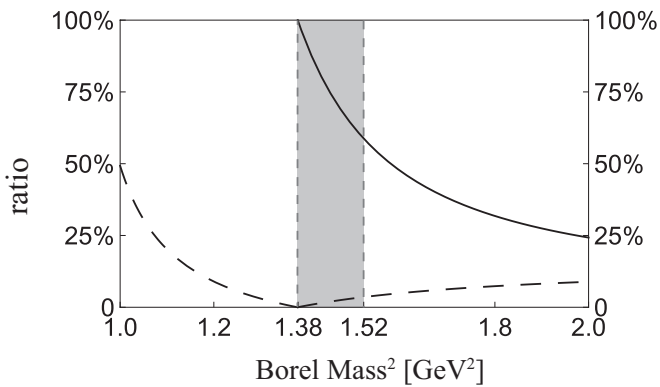}
\end{center}
\caption{Off-diagonal terms, $\left|\Pi_{12}^2/(\Pi_{11}\Pi_{22})\right|$ (solid) and $\left|(\Pi_{J_1J_2})^2/(\Pi_{J_1J_1}\Pi_{J_2J_2})\right|$ (dashed), as functions of the Borel mass $M_B$. These curves are obtained by setting $s_0 = 7.6$~GeV$^2$.
}
\label{fig:offdiagonal}
\end{figure}

Firstly, we examine how large is the off-diagonal term $\Pi_{12}(M_B^2)$ defined in Eq.~(\ref{eq:offdiag1}). As shown in Fig.~\ref{fig:offdiagonal} using the solid curve, the ratio $\Pi_{12}^2/(\Pi_{11}\Pi_{22})$ is quite large, so the mixing should be taken into account. Accordingly, we diagonalize the matrix
\begin{eqnarray}
\left( \begin{array}{cc}
\Pi_{11}(s_0,M_B^2) & \Pi_{12}(s_0,M_B^2)
\\ \Pi_{12}^\dagger(s_0,M_B^2) & \Pi_{22}(s_0,M_B^2)
\end{array} \right) \, .
\end{eqnarray}
at around $M_B^2 = 1.4$~GeV$^2$ and $s_0 = 7.6$~GeV$^2$ (we shall see that these two values are both inside the working regions for the mixed current $J_{2}$). We obtain two new currents with the mixing angle $\theta = 16.3^{\rm o}$:
\begin{eqnarray}
J_{1} &=& \cos\theta~\eta_{1} + \sin\theta~\eta_{2} \, ,
\\ \nonumber J_{2} &=& - \sin\theta~\eta_{1} + \cos\theta~\eta_{2} \, ,
\end{eqnarray}
As shown in Fig.~\ref{fig:offdiagonal} using the dashed curve, the new ratio $(\Pi_{J_1J_2})^2/(\Pi_{J_1J_1}\Pi_{J_2J_2})$ is significantly suppressed in the region $1.38$~GeV$^2 < M_B^2 < 1.52$~GeV$^2$ (Borel window for $J_{2}$ when setting $s_0 = 7.6$~GeV$^2$), so $J_{1}$ and $J_{2}$ only weakly correlate with each other inside this region.

We separately use $J_{1}$ and $J_{2}$ to perform QCD sum rule analyses. When using $J_1$, we find that there exist non-vanishing Borel windows as long as $s_0 \geq 9.4$~GeV$^2$, and the mass extracted is around 3.14~GeV, even larger than 3.0~GeV, so we shall not use it to draw any conclusion.

\begin{figure*}[]
\begin{center}
\includegraphics[width=0.45\textwidth]{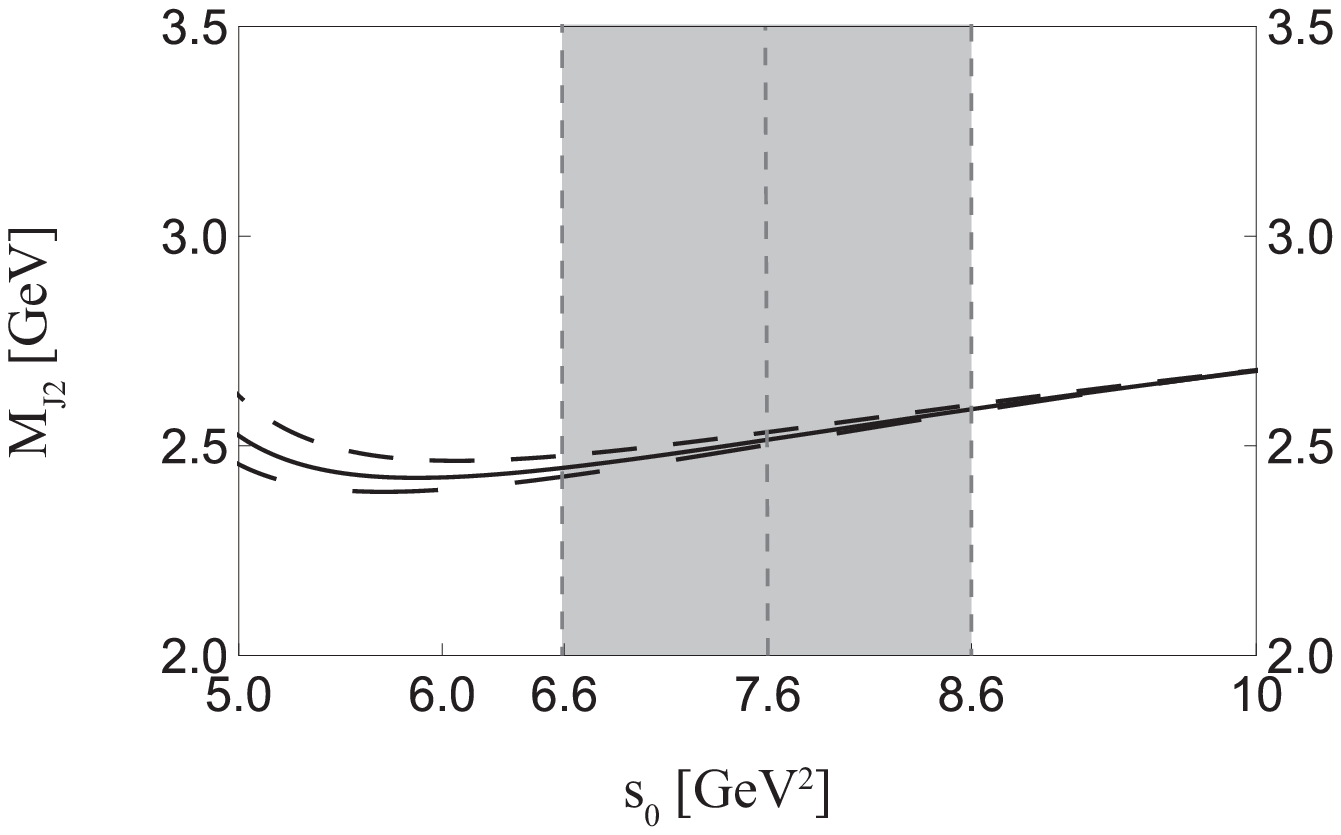}
~~~~~
\includegraphics[width=0.45\textwidth]{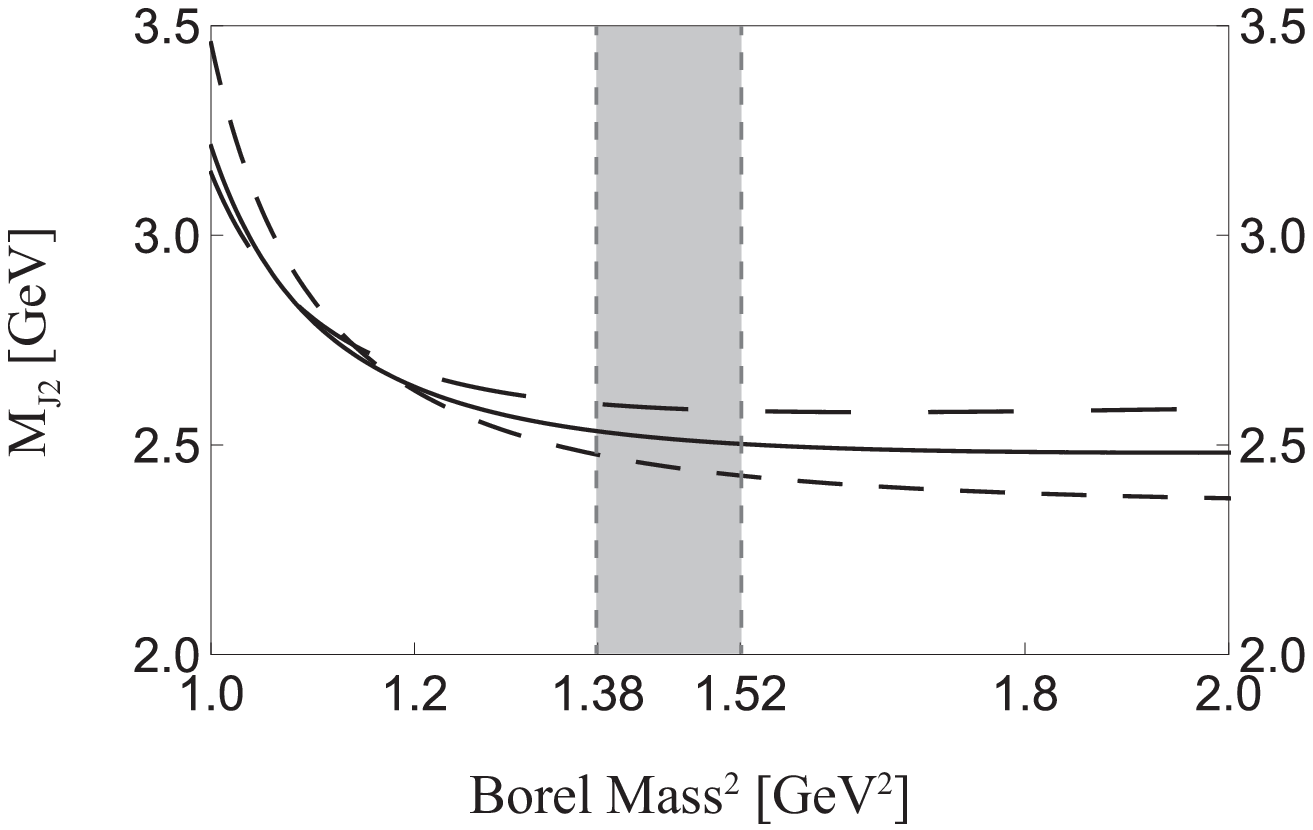}
\caption{
Mass calculated using the current $J_2$, as a function of the threshold value $s_0$ (left) and the Borel mass $M_B$ (right).
In the left panel the short-dashed/solid/long-dashed curves are obtained by setting $M_B^2 = 1.38/1.45/1.52$ GeV$^2$, respectively.
In the right panel the short-dashed/solid/long-dashed/dotted curves are obtained by setting $s_0 = 6.6/7.6/8.6$ GeV$^2$, respectively.}
\label{fig:J2mass}
\end{center}
\end{figure*}

When using $J_{2}$, we find that there exist non-vanishing Borel windows as long as $s_0 \geq 7.2$~GeV$^2$. We show the mass extracted from $J_2$ in Fig.~\ref{fig:J2mass} as a function of the threshold value $s_0$ (left) and the Borel mass $M_B$ (right). Using the working regions $6.6$~GeV$^2< s_0 < 8.6$~GeV$^2$ and $1.38$~GeV$^2 < M_B^2 < 1.52$~GeV$^2$ (Borel window for $s_0=7.6$~GeV$^2$), we obtain
\begin{equation}
M_{J_2} = 2.51^{+0.15}_{-0.12} {\rm~GeV} \, ,
\end{equation}
where the central value corresponds to $s_0=7.6$~GeV$^2$ and $M_B^2 = 1.45$~GeV$^2$. Here we have temporarily assumed the uncertainty of the mixing angle to be $\theta = 16.3^{\rm o} \pm 10.0^{\rm o}$, since $J_{1}$ and $J_{2}$ become correlated again when $\theta$ is outside this region. The mass uncertainty due to this angle is $2.51^{+0.04}_{-0.03}$~GeV, that is not so large.

%
%

%
\section{Summary and Discussions}
\label{sec:summary}
%

In this paper we use the method of QCD sum rules to study the $ss\bar s\bar s$ tetraquark states of $J^{PC} = 0^{-+}$. We systematically construct all the relevant diquark-antidiquark $(ss)(\bar s \bar s)$ and meson-meson $(\bar s s)(\bar s s)$ interpolating currents, and derive their relations through the Fierz transformation. We find two independent currents $\eta_1$ and $\eta_2$, and calculate both their diagonal and off-diagonal two-point correlation functions. The obtained results suggest that these two single currents strongly correlate with each other. Hence, we use them to further construct two mixed currents $J_{1}$ and $J_{2}$, which only weakly correlate with each other.

We use the two single currents $\eta_1$ and $\eta_2$ as well as the two mixed currents $J_{1}$ and $J_{2}$ to perform QCD sum rule analyses. We find the correlation functions $\Pi_{11}(M_B^2)$ and $\Pi_{22}(M_B^2)$ to be both negative in the region $s_0 < 4.0$ GeV$^2$ when taking $M_B^2 = 1.5$~GeV$^2$. This suggests that these currents couple weakly to the lower state $\eta(1475)$ as well as the $\eta^\prime f_0(980)$ threshold, so the state they couple to, as if they can couple to some state, should be new and possibly exotic.

After performing numerical analyses, we extract the masses from $\eta_1$ and $\eta_2$ to be $2.86^{+0.18}_{-0.12}$~GeV and $2.59^{+0.14}_{-0.10}$~GeV respectively, and the masses from $J_{1}$ and $J_{2}$ to be around 3.14~GeV and $2.51^{+0.14}_{-0.12}$~GeV respectively. These mass values are not affected much by the lower state $\eta(1475)$ as well as the $\eta^\prime f_0(980)$ threshold, because the currents couple weakly to them. However, there may exist some other thresholds, which are difficult to be fully taken into account.

Especially, the mass extracted from the mixed current $J_{2}$ is the lowest:
\begin{equation}
M_{J_2} = 2.51 ^{+0.15}_{-0.12}  {\rm~GeV} \, .
\end{equation}
Use the Fierz transformation given in Eqs.~(\ref{eq:fierz2}), we can transform $J_{2}$ to be
\begin{eqnarray}
J_{2} &=& - \sin16.3^{\rm o}~\eta_{1} + \cos16.3^{\rm o}~\eta_{2}
\\ \nonumber &=& 6.04 ~ \eta_3 - 0.55 ~ \eta_4
\\ \nonumber &=& 6.04 ~ (\bar{s}_a s_a)(\bar{s}_b \gamma_5 s_b)
- 0.55 ~ (\bar{s}_a \sigma_{\mu\nu} s_a)(\bar{s}_b \sigma^{\mu\nu} \gamma_5 s_b) \, .
\end{eqnarray}
This suggests that the state $X$, coupled by this current, can decay into the following channels:
\begin{itemize}

\item It can decay into the $\eta^\prime f_0(980)$ channel, due to the $(\bar{s}_a s_a)(\bar{s}_b \gamma_5 s_b)$ operator~\cite{Chen:2019wjd,Cheng:2005nb}:
\begin{eqnarray}
\langle 0 | \bar{s}_a s_a | f_0(980) \rangle &=& f_{f_0(980)} m_{f_0(980)} \, ,
\\ \langle 0 | \bar{s}_b i\gamma_5 s_b | \eta^\prime \rangle &=& \lambda_{\eta^\prime} \, ,
\end{eqnarray}
where $f_{f_0(980)}$ and $\lambda_{\eta^\prime}$ are decay constants. Considering that the $f_0(980)$ resonance can further decay into the $\pi\pi$ and $K \bar K$ final states, we use the BaBar measurement~\cite{Aubert:2006nu}:
    \begin{equation}
    {\mathcal{B}({f_0(980) \to K^+ K^-}) \over \mathcal{B}({f_0(980) \to \pi^+ \pi^-})} = 0.69 \pm 0.32 \, ,
    \end{equation}
    to further estimate and obtain
    \begin{equation}
    {\mathcal{B}({X \to \eta^\prime f_0(980) \to \eta^\prime K \bar K}) \over \mathcal{B}({X \to \eta^\prime f_0(980) \to \eta^\prime \pi \pi})} = 0.92 \pm 0.43 \, .
    \end{equation}

\item It can also decay into the $\phi \phi$ final state, due to the $(\bar{s}_a \sigma_{\mu\nu} s_a)(\bar{s}_b \sigma^{\mu\nu} \gamma_5 s_b)$ operator:
\begin{eqnarray}
\langle 0 | \bar{s}_a \sigma_{\mu\nu} s_a | \phi(p,\epsilon) \rangle &=& i f_{\phi}^T (p_\mu \epsilon_\nu - p_\nu \epsilon_\mu) \, ,
\\ \langle 0 | \bar{s}_a \sigma_{\mu\nu} \gamma_5 s_a | \phi(p,\epsilon) \rangle &=& - f_{\phi}^T \epsilon_{\mu\nu\rho\sigma} p^\rho \epsilon^\sigma \, ,
\end{eqnarray}
where $f_{\phi}^T$ is the decay constant.

\end{itemize}
In the three BESIII experiments~\cite{Ablikim:2010au,Ablikim:2019zyw,Ablikim:2016hlu}, the $X(2370)$ was observed in both the $\eta^\prime \pi\pi$ and $\eta^\prime K \bar K$ final states, and the $X(2500)$ was observed in the $\phi \phi$ final state, indicating that both of them contain many $strangeness$ components. Accordingly, our results suggest that the $X(2500)$ can be well explained as the $ss\bar s\bar s$ tetraquark state of $J^{PC} = 0^{-+}$, and the $X(2370)$ may also be explained as such a state (they might even be the same state, so that its mass spectrum and decay properties can both be well explained).
To verify the above interpretation, we propose the BESIII Collaboration to further study the $\pi\pi$ and $K \bar K$ invariant mass spectra of the $J/\psi \to \gamma \pi \pi \eta^\prime$ and $J/\psi \to \gamma K \bar K \eta^\prime$ decays to examine whether there exists the $f_0(980)$ resonance.

%
\section*{Acknowledgments}
%

We thank Cheng-Ping Shen for useful discussions.
This project is supported by
the National Natural Science Foundation of China under Grant No.~11722540
and
the Fundamental Research Funds for the Central Universities.


\begin{thebibliography}{99}

\bibitem{pdg}
  C.~Patrignani {\it et al.} [Particle Data Group],
  {\it Review of Particle Physics},
  \href{http://dx.doi.org/10.1088/1674-1137/40/10/100001}{Chin.\ Phys.\ C {\bf 40}, 100001 (2016)}.

\bibitem{Liu:2019zoy}
  Y.~R.~Liu, H.~X.~Chen, W.~Chen, X.~Liu and S.~L.~Zhu,
  {\it Pentaquark and Tetraquark States},
  \href{http://dx.doi.org/10.1016/j.ppnp.2019.04.003}{Prog.\ Part.\ Nucl.\ Phys.\  {\bf 107}, 237 (2019)}.

\bibitem{Lebed:2016hpi}
  R.~F.~Lebed, R.~E.~Mitchell and E.~S.~Swanson,
  {\it Heavy-quark QCD exotica},
  \href{http://dx.doi.org/10.1016/j.ppnp.2016.11.003}{Prog.\ Part.\ Nucl.\ Phys.\  {\bf 93}, 143 (2017)}.

\bibitem{Esposito:2016noz}
  A.~Esposito, A.~Pilloni and A.~D.~Polosa,
  {\it Multiquark resonances},
  \href{http://dx.doi.org/10.1016/j.physrep.2016.11.002}{Phys.\ Rept.\  {\bf 668}, 1 (2017)}.

\bibitem{Guo:2017jvc}
  F.~K.~Guo, C.~Hanhart, U.~G.~Mei{\ss}ner, Q.~Wang, Q.~Zhao, and B.~S.~Zou,
  {\it Hadronic molecules},
  \href{http://dx.doi.org/10.1103/RevModPhys.90.015004}{Rev.\ Mod.\ Phys.\  {\bf 90}, 015004 (2018)}.

\bibitem{Ali:2017jda}
  A.~Ali, J.~S.~Lange and S.~Stone,
  {\it Exotics: Heavy pentaquarks and tetraquarks},
  \href{http://dx.doi.org/10.1016/j.ppnp.2017.08.003}{Prog.\ Part.\ Nucl.\ Phys.\  {\bf 97}, 123 (2017)}.

\bibitem{Olsen:2017bmm}
  S.~L.~Olsen, T.~Skwarnicki, and D.~Zieminska,
  {\it Nonstandard heavy mesons and baryons: Experimental evidence},
  \href{http://dx.doi.org/10.1103/RevModPhys.90.015003}{Rev.\ Mod.\ Phys.\  {\bf 90}, 015003 (2018)}.

\bibitem{Karliner:2017qhf}
  M.~Karliner, J.~L.~Rosner and T.~Skwarnicki,
  {\it Multiquark States},
  \href{http://dx.doi.org/10.1146/annurev-nucl-101917-020902}{Ann.\ Rev.\ Nucl.\ Part.\ Sci.\  {\bf 68}, 17 (2018)}.

\bibitem{Brambilla:2019esw}
  N.~Brambilla, S.~Eidelman, C.~Hanhart, A.~Nefediev, C.~P.~Shen, C.~E.~Thomas, A.~Vairo and C.~Z.~Yuan,
  {\it The $XYZ$ states: experimental and theoretical status and perspectives},
  \href{http://arxiv.org/abs/arXiv:1907.07583}{arXiv:1907.07583 [hep-ex]}.

\bibitem{Guo:2019twa}
  F.~K.~Guo, X.~H.~Liu and S.~Sakai,
  {\it Threshold cusps and triangle singularities in hadronic reactions},
  \href{http://arxiv.org/abs/arXiv:1912.07030}{arXiv:1912.07030 [hep-ph]}.

\bibitem{Ablikim:2010au}
  M.~Ablikim {\it et al.} [BESIII Collaboration],
  {\it Confirmation of the $X(1835)$ and observation of the resonances $X(2120)$ and $X(2370)$ in $J/\psi\to \gamma \pi^+\pi^-\eta^\prime$},
   \href{http://dx.doi.org/10.1103/PhysRevLett.106.072002}{Phys.\ Rev.\ Lett.\  {\bf 106}, 072002 (2011)}.

\bibitem{Bai:2003sw}
  J.~Z.~Bai {\it et al.} [BES Collaboration],
  {\it Observation of a near threshold enhancement in th p anti-p mass spectrum from radiative J / psi ---> gamma p anti-p decays},
  \href{http://dx.doi.org/10.1103/PhysRevLett.91.022001}{Phys.\ Rev.\ Lett.\  {\bf 91}, 022001 (2003)}.

\bibitem{Ablikim:2005um}
  M.~Ablikim {\it et al.} [BES Collaboration],
  {\it Observation of a resonance $X(1835)$ in $J/\psi \to \gamma \pi^+ \pi^- \eta^\prime$},
  \href{http://dx.doi.org/10.1103/PhysRevLett.95.262001}{Phys.\ Rev.\ Lett.\  {\bf 95}, 262001 (2005)}.

\bibitem{BESIII:2010krt}
  M.~Ablikim {\it et al.} [BESIII Collaboration],
  {\it Observation of a $p\bar{p}$ mass threshoud enhancement in $\psi^\prime\to\pi^+\pi^-J/\psi(J/\psi\to\gamma p\bar{p})$ decay},
  \href{http://dx.doi.org/10.1088/1674-1137/34/4/001}{Chin.\ Phys.\ C {\bf 34}, 421}.

\bibitem{Ablikim:2019zyw}
  M.~Ablikim {\it et al.} [BESIII Collaboration],
  {\it Observation of the $X(2370)$ and Search for the $X(2120)$ in $J/\psi\to\gamma K\bar{K} \eta'$},
  \href{http://arxiv.org/abs/1912.11253}{arXiv:1912.11253 [hep-ex]}.

\bibitem{Ablikim:2016hlu}
  M.~Ablikim {\it et al.} [BESIII Collaboration],
  {\it Observation of pseudoscalar and tensor resonances in $J/\psi\to \gamma \phi \phi$},
  \href{http://dx.doi.org/10.1103/PhysRevD.93.112011}{Phys.\ Rev.\ D {\bf 93}, 112011 (2016)}.

\bibitem{Ablikim:2020pgw}
  M.~Ablikim {\it et al.},
  {\it Observation of a resonant structure in $e^{+}e^{-} \to K^{+}K^{-}\pi^{0}\pi^{0}$},
  \href{http://arxiv.org/abs/2001.04131}{arXiv:2001.04131 [hep-ex]}.


\bibitem{Kou:2018nap}
  E.~Kou {\it et al.} [Belle-II Collaboration],
  {\it The Belle II Physics Book},
  \href{http://dx.doi.org/10.1093/ptep/ptz106}{PTEP {\bf 2019}, 123C01 (2019)}.

\bibitem{Austregesilo:2018mno}
  A.~Austregesilo [GlueX Collaboration],
  {\it Light-Meson Spectroscopy at GlueX},
  \href{http://dx.doi.org/10.1142/S2010194518600297}{Int.\ J.\ Mod.\ Phys.\ Conf.\ Ser.\  {\bf 46}, 1860029 (2018)}.

\bibitem{Chen:2008ej}
  H.~X.~Chen, X.~Liu, A.~Hosaka and S.~L.~Zhu,
  {\it The Y(2175) State in the QCD Sum Rule},
  \href{http://dx.doi.org/10.1103/PhysRevD.78.034012}{Phys.\ Rev.\ D {\bf 78}, 034012 (2008)}.

\bibitem{Chen:2018kuu}
  H.~X.~Chen, C.~P.~Shen and S.~L.~Zhu,
  {\it A possible partner state of the $Y(2175)$},
  \href{http://dx.doi.org/10.1103/PhysRevD.98.014011}{Phys.\ Rev.\ D {\bf 98}, 014011 (2018)}.

\bibitem{Cui:2019roq}
  E.~L.~Cui, H.~M.~Yang, H.~X.~Chen, W.~Chen and C.~P.~Shen,
  {\it QCD sum rule studies of $s s {\bar{s}} {\bar{s}}$ tetraquark states with $J^{PC} = 1^{+-}$},
  \href{http://dx.doi.org/10.1140/epjc/s10052-019-6755-y}{Eur.\ Phys.\ J.\ C {\bf 79}, 232 (2019)}.

\bibitem{Liu:2010tr}
  J.~F.~Liu {\it et al.} [BES Collaboration],
  {\it $X(1835)$ and the New Resonances $X(2120)$ and $X(2370)$ Observed by the BES Collaboration},
  \href{http://dx.doi.org/10.1103/PhysRevD.82.074026}{Phys.\ Rev.\ D {\bf 82}, 074026 (2010)}.

\bibitem{Qin:2017qes}
  W.~Qin, Q.~Zhao and X.~H.~Zhong,
  {\it Revisiting the pseudoscalar meson and glueball mixing and key issues in the search for a pseudoscalar glueball state},
  \href{http://dx.doi.org/10.1103/PhysRevD.97.096002}{Phys.\ Rev.\ D {\bf 97}, 096002 (2018)}.



\bibitem{Yu:2011ta}
  J.~S.~Yu, Z.~F.~Sun, X.~Liu and Q.~Zhao,
  {\it Categorizing resonances X(1835), X(2120) and X(2370) in the pseudoscalar meson family},
  \href{http://dx.doi.org/10.1103/PhysRevD.83.114007}{Phys.\ Rev.\ D {\bf 83}, 114007 (2011)}.

\bibitem{Deng:2012wi}
  C.~Deng, J.~Ping, Y.~Yang and F.~Wang,
  {\it X(1835), X(2120) and X(2370) in a flux tube model},
  \href{http://dx.doi.org/10.1103/PhysRevD.86.014008}{Phys.\ Rev.\ D {\bf 86}, 014008 (2012)}.

\bibitem{Eshraim:2012jv}
  W.~I.~Eshraim, S.~Janowski, F.~Giacosa and D.~H.~Rischke,
  {\it Decay of the pseudoscalar glueball into scalar and pseudoscalar mesons},
  \href{http://dx.doi.org/10.1103/PhysRevD.87.054036}{Phys.\ Rev.\ D {\bf 87}, 054036 (2013)}.

\bibitem{Gui:2019dtm}
  L.~C.~Gui, J.~M.~Dong, Y.~Chen and Y.~B.~Yang,
  {\it Study of the pseudoscalar glueball in $J/\psi$ radiative decays},
  \href{http://dx.doi.org/10.1103/PhysRevD.100.054511}{Phys.\ Rev.\ D {\bf 100}, 054511 (2019)}.



\bibitem{Pan:2016bac}
  T.~T.~Pan, Q.~F.~L¨¹, E.~Wang and D.~M.~Li,
  {\it Strong decays of the $X(2500)$ newly observed by the BESIII Collaboration},
  \href{http://dx.doi.org/10.1103/PhysRevD.94.054030}{Phys.\ Rev.\ D {\bf 94}, 054030 (2016)}.

\bibitem{Xue:2018jvi}
  S.~C.~Xue, G.~Y.~Wang, G.~N.~Li, E.~Wang and D.~M.~Li,
  {\it The possible members of the $5^1S_0$ meson nonet},
  \href{http://dx.doi.org/10.1140/epjc/s10052-018-5961-3}{Eur.\ Phys.\ J.\ C {\bf 78}, 479 (2018)}.


\bibitem{Wang:2017iai}
  L.~M.~Wang, S.~Q.~Luo, Z.~F.~Sun and X.~Liu,
  {\it Constructing new pseudoscalar meson nonets with the observed $X(2100)$, $X(2500)$, and $\eta(2225)$},
  \href{http://dx.doi.org/10.1103/PhysRevD.96.034013}{Phys.\ Rev.\ D {\bf 96}, 034013 (2017)}.

\bibitem{Morningstar:1999rf}
  C.~J.~Morningstar and M.~J.~Peardon,
  {\it The Glueball spectrum from an anisotropic lattice study},
  \href{http://dx.doi.org/10.1103/PhysRevD.60.034509}{Phys.\ Rev.\ D {\bf 60}, 034509 (1999)}.

\bibitem{Chen:2005mg}
  Y.~Chen {\it et al.},
  {\it Glueball spectrum and matrix elements on anisotropic lattices},
  \href{http://dx.doi.org/10.1103/PhysRevD.73.014516}{Phys.\ Rev.\ D {\bf 73}, 014516 (2006)}.

\bibitem{Richards:2010ck}
  C.~M.~Richards {\it et al.} [UKQCD Collaboration],
  {\it Glueball mass measurements from improved staggered fermion simulations},
   \href{http://dx.doi.org/10.1103/PhysRevD.82.034501}{Phys.\ Rev.\ D {\bf 82}, 034501 (2010)}.

\bibitem{Gregory:2012hu}
  E.~Gregory, A.~Irving, B.~Lucini, C.~McNeile, A.~Rago, C.~Richards and E.~Rinaldi,
  {\it Towards the glueball spectrum from unquenched lattice QCD},
   \href{http://dx.doi.org/10.1007/JHEP10(2012)170}{JHEP {\bf 1210}, 170 (2012)}.

\bibitem{Eshraim:2016mds}
  W.~I.~Eshraim and S.~Schramm,
  {\it Decay modes of the excited pseudoscalar glueball},
  \href{http://dx.doi.org/10.1103/PhysRevD.95.014028}{Phys.\ Rev.\ D {\bf 95}, 014028 (2017)}.

\bibitem{Eshraim:2019sgr}
  W.~I.~Eshraim,
  {\it Decay of the pseudoscalar glueball and its first excited state into scalar and pseudoscalar mesons and their first excited states},
  \href{http://dx.doi.org/10.1103/PhysRevD.100.096007}{Phys.\ Rev.\ D {\bf 100}, 096007 (2019)}.

\bibitem{Napsuciale:2007wp}
  M.~Napsuciale, E.~Oset, K.~Sasaki and C.~A.~Vaquera-Araujo,
  \href{http://dx.doi.org/10.1103/PhysRevD.76.074012}{Phys.\ Rev.\ D {\bf 76}, 074012 (2007)}.

\bibitem{MartinezTorres:2008gy}
  A.~Martinez Torres, K.~P.~Khemchandani, L.~S.~Geng, M.~Napsuciale and E.~Oset,
  \href{http://dx.doi.org/10.1103/PhysRevD.78.074031}{Phys.\ Rev.\ D {\bf 78}, 074031 (2008)}.

\bibitem{Liang:2013yta}
  W.~Liang, C.~W.~Xiao and E.~Oset,
  {\it Study of $\eta K\overline{K}$ and $\eta^\prime K \overline{K}$ with the fixed center approximation to Faddeev equations},
  \href{http://dx.doi.org/10.1103/PhysRevD.88.114024}{Phys.\ Rev.\ D {\bf 88}, 114024 (2013)}.

\bibitem{Kozhevnikov:2019lmy}
  A.~A.~Kozhevnikov,
  {\it Dynamical analysis of the $X$ resonance contributions to the decay $J/\psi\to\gamma X\to\gamma\phi\phi$},
   \href{http://dx.doi.org/10.1103/PhysRevD.99.014019}{Phys.\ Rev.\ D {\bf 99}, 014019 (2019)}.

\bibitem{Lebiedowicz:2019jru}
  P.~Lebiedowicz, O.~Nachtmann and A.~Szczurek,
  {\it Central exclusive diffractive production of $K^{+} K^{-} K^{+} K^{-}$ via the intermediate $\phi \phi$ state in proton-proton},
   \href{http://dx.doiorg/10.1103/PhysRevD.99.094034}{Phys.\ Rev.\ D {\bf 99}, 094034 (2019)}.

\bibitem{Kozhevnikov:2019rma}
  A.~A.~Kozhevnikov,
  {\it The decay $J/ \psi\rightarrow\gamma\phi\phi$ : Spin dependence of amplitude and angular distributions of photons with linear polarizations},
  \href{http://dx.doi.org/10.1140/epja/i2019-12845-8}{Eur.\ Phys.\ J.\ A {\bf 55}, 155 (2019)}.

\bibitem{Shifman:1978bx}
  M.~A.~Shifman, A.~I.~Vainshtein and V.~I.~Zakharov,
  {\it QCD And Resonance Physics. Sum Rules},
  \href{http://dx.doi.org/10.1016/0550-3213(79)90022-1}{Nucl.\ Phys.\  {\bf B  147}, 385 (1979)}.

\bibitem{Reinders:1984sr}
  L.~J.~Reinders, H.~Rubinstein and S.~Yazaki,
  {\it Hadron Properties From QCD Sum Rules},
  \href{http://dx.doi.org/10.1016/0370-1573(85)90065-1}{Phys.\ Rept.\  {\bf 127}, 1 (1985)}.

\bibitem{Yang:1993bp}
  K.~C.~Yang, W.~Y.~P.~Hwang, E.~M.~Henley and L.~S.~Kisslinger,
  {\it QCD sum rules and neutron proton mass difference},
  \href{http://dx.doi.org/10.1103/PhysRevD.47.3001}{Phys.\ Rev.\ {\bf D  47}, 3001 (1993)}.

\bibitem{Narison:2002pw}
  S.~Narison,
  {\it QCD as a theory of hadrons (from partons to confinement)},
  Camb.\ Monogr.\ Part.\ Phys.\ Nucl.\ Phys.\ Cosmol.\  {\bf 17}, 1 (2002).

\bibitem{Gimenez:2005nt}
  V.~Gimenez, V.~Lubicz, F.~Mescia, V.~Porretti and J.~Reyes,
  {\it Operator product expansion and quark condensate from lattice QCD in
  coordinate space},
  \href{http://dx.doi.org/10.1140/epjc/s2005-02250-9}{Eur.\ Phys.\ J.\ {\bf C  41}, 535 (2005)}.

\bibitem{Jamin:2002ev}
  M.~Jamin,
  {\it Flavour-symmetry breaking of the quark condensate and chiral  corrections
  to the Gell-Mann-Oakes-Renner relation},
  \href{http://dx.doi.org/10.1016/S0370-2693(02)01951-2}{Phys.\ Lett.\ {\bf B  538}, 71 (2002)}.

\bibitem{Ioffe:2002be}
  B.~L.~Ioffe and K.~N.~Zyablyuk,
  {\it Gluon condensate in charmonium sum rules with 3-loop corrections},
  \href{http://dx.doi.org/10.1140/epjc/s2002-01099-8}{Eur.\ Phys.\ J.\ {\bf C  27}, 229 (2003)}.

\bibitem{Ovchinnikov:1988gk}
  A.~A.~Ovchinnikov and A.~A.~Pivovarov,
  {\it QCD Sum Rule Calculation Of The Quark Gluon Condensate},
  Sov.\ J.\ Nucl.\ Phys.\  {\bf 48}, 721 (1988)
  [Yad.\ Fiz.\  {\bf 48}, 1135 (1988)].

\bibitem{Ellis:1996xc}
  J.~R.~Ellis, E.~Gardi, M.~Karliner and M.~A.~Samuel,
  {\it Renormalization-scheme dependence of Pade summation in QCD},
  \href{http://dx.doi.org/10.1103/PhysRevD.54.6986}{Phys.\ Rev.\  {\bf D  54}, 6986 (1996)}.

\bibitem{Cheng:2005nb}
  H.~Y.~Cheng, C.~K.~Chua and K.~C.~Yang,
  {\it Charmless hadronic B decays involving scalar mesons: Implications to the nature of light scalar mesons},
  \href{http://dx.doi.org/10.1103/PhysRevD.73.014017}{Phys.\ Rev.\ D {\bf 73}, 014017 (2006)}.

\bibitem{Chen:2019wjd}
  H.~X.~Chen,
  {\it Decay properties of the $Z_c(3900)$ through the Fierz rearrangement},
  \href{http://arxiv.org/abs/1910.03269}{arXiv:1910.03269 [hep-ph]}.

\bibitem{Aubert:2006nu}
  B.~Aubert {\it et al.} [BaBar Collaboration],
  {\it Dalitz plot analysis of the decay $B^\pm \to K^\pm K^\pm K^\mp$},
  \href{http://dx.doi.org/10.1103/PhysRevD.74.032003}{Phys.\ Rev.\ D {\bf 74}, 032003 (2006)}.

\end{thebibliography}
\end{document}